\documentclass[a4paper,12pt,english,german]{article}

\usepackage{mathtools}
\usepackage{graphicx}

\begin{document}

\centerline{\bf  Violation of the Born Rule: Implications for Macroscopic Fields.}
\bigskip

\centerline{ R. E. Kastner\footnote{Foundations of Physics Group, University of Maryland, College
Park, USA;  rkastner@umd.edu}}
 \centerline{20 June 2016}

\bigskip

{\small ABSTRACT. It is shown that violation of the Born Rule leads to a breakdown of the correspondence between the quantum electromagnetic field and its classical counterpart. Specifically, the relationship of the quantum coherent state to the  classical electromagnetic field turns out to imply that if the Born Rule were violated, this could result in apparent deviations from the energy conservation law applying to the field and its sources (Poynting's Theorem). The result, which is fully general and independent of interpretations of quantum theory, suggests that the Born Rule is just as fundamental a law of Nature as are the field conservation laws.}

\bigskip

\section{Introduction}

The Born Rule, which states that the probability of outcome $x$ for a quantum described by a wavefunction $\Psi(x)$ is the absolute square of the wavefunction, was famously first proposed as an educated guess by Max Born. Born included it as a note added in proof correcting his original supposition that the wavefunction $\Psi(x)$ itself yielded the probability for outcome $x$.\cite{Born}. Since then, the Born Rule has been assumed as a basic axiom of quantum theory (although there are attempts to derive the rule from operationalist and/or decision-theoretic assumptions, e.g. \cite{Saunders}). While a physical derivation is available in the transactional interpretation\cite{Cramer},\cite{Kastner}, it should be emphasized that the arguments herein do not depend on any particular interpretation of quantum theory. The demonstrated implications for violations of the Born Rule hold regardless of one's approach to deriving or explaining the rule. Indeed they hold if the rule is taken as an unexplained axiom.

Since the implications explored herein affect classical field theories, it should also be noted that such theories are widely understood to be only limiting or approximate accounts of Nature. For example, it was the failure of classical electromagnetism to account for the stability of atoms that led to the advent of quantum theory in the first place. When we use a classical description, we are using an approximation or idealization that is useful and approximately true under the conditions obtaining. So, as noted for example by Sakurai,\cite{Sakurai} (p. 38), the classical description of the electromagnetic field obtains in the limit in which the number $n$ of photons in the field is so large that it approaches a continuous parameter. For $n$ decreasing from that limit, the classical field description becomes less and less accurate until it fails completely. Thus, at bottom, the field is quantum mechanical. The classical field theory turns out to be an idealization, even if an important and useful one.

In what follows, we examine a particular kind of quantum state that underlies the (apparently) classical electromagnetic field--the coherent state or Glauber state--and explore what happens to the classical field equations if the probabilities associated with that state deviate from the Born Rule.

\section{Coherent states}

Coherent states, often called Glauber states due to the pioneering work of Roy Glauber\cite{Glauber}, are the quantum states that support classical electromagnetic fields. Whereas so-called Fock states $| n\rangle$ are eigenstates of the occupation
number operator $\hat{N}$ with definite photon number $n$, coherent states are states for which the quantum electromagnetic field has an indefinite number of photons. 

The coherent state $|\alpha\rangle$  is defined in terms of Fock states as

$$ |\alpha\rangle = e^{-\frac{|\alpha |^2}{2} }  \sum_{n=0}^{\infty}  \frac{\alpha^n}{\sqrt{n!}} | n \rangle      \eqno(1) $$

The coherent state $|\alpha\rangle$ is an eigenstate of the field annihilation operator $\hat{a}$ with eigenvalue $\alpha$ (which can be complex, since $\hat{a}$ is not Hermitian). Thus we have that

$$ \hat{a} | \alpha\rangle = \alpha  | \alpha\rangle $$  and

$$  \langle  \alpha | \hat{a}^\dag = \langle \alpha | \alpha^*  \eqno(2a,b)$$ 
 
 The probability of detecting a particular number $n$ in the state $|\alpha\rangle $ is given by the Born Rule, i.e.,
 
 $$P(n) = |\langle n |\alpha\rangle|^2 = e^{-\langle n \rangle} \frac{\langle n \rangle^n}{n!} \eqno(3) $$

\noindent It is a Poisson distribution, reflecting the statistical independence of each detection.

The mean photon number $\langle n \rangle$ is given by the expectation value of the number operator 
$\hat{N}= \hat{a}^\dag \hat{a}$ for $ | \alpha\rangle $, namely:
 
 $$ \langle n \rangle = \langle  \hat{a}^\dag \hat{a} \rangle =   |\alpha |^2 \eqno(4) $$
  
 \section{Correspondence between the classical electromagnetic field and the quantum field}
  
  The electric field $\hat{E}$ (which is an operator in the quantum electromagnetic field) is proportional to the ``field quadrature'' $X$,
  
  $$ X = \hat{a}^\dag + \hat{a} \eqno (5) $$
  
  Upon taking the expectation value of $X$  for the coherent state $| \alpha \rangle$ (assuming for simplicity only a single field mode), we find based on (2) that 
  
  $$ \langle X \rangle = \alpha^* + \alpha  \eqno(6) $$
  
 \noindent  i.e. the coherent state corresponds to a classical electic field with amplitude $\alpha $.
  \smallskip
  
  It is well-known that the classical electromagnetic field obeys energy conservation; this is usually stated in terms of Poynting's Theorem. The ``Poynting Vector'' expressing energy flow is denoted ${\bf S}$. For a collection of fields and sources in some arbitrary volume V bounded by surface $\sigma$,
  
  $$ \frac{dE}{dt} = \frac{d}{dt} (E_{mech} + E_{field}) = -\oint_{\sigma} \bf{n} \cdot \bf{S} \ da \eqno(7) $$
  
  \noindent where $E_{field}$ is the standard classical field energy, 
  
  $$ E_{field} = \frac{1}{8\pi} \int (\bf{E}^2 + \bf{B}^2 ) d^3 x \eqno(8) $$
  
  It should be noted that (8) is a straightforward implication of Maxwell's Equations (see, e.g. \cite{Reitz}, \textsection 16-3).
  
In (7),  $E_{mech}$ is the rate of work done on all charges in V,
  
    $$ \frac{d E_{mech}}{dt}  = \int \bf{J} \cdot \bf{E} \ d^3 x \eqno (9) $$
    
 \noindent   and the Poynting vector ${\bf S} $ is explicitly given by
    
    $$\bf{S} = \frac{c}{4\pi} (\bf{E} \times \bf{H} )  \eqno(10). $$
  
  An ordinary classical field is instantiated by a coherent state with an enormous average photon number. However, much smaller amplitude fields can be studied in the laboratory. As the average photon number increases, the more closely the measured field approaches the classical ideal\cite{Breitechbach}. It is important to keep in mind that at a fundamental level all field measurements are achieved through photon absorption; i.e., absorption of electromagnetic energy, which is proportional to the square of the field amplitude. One never detects the amplitude directly but always infers it based on the energy transferred by the field. Thus, it is the average photon number $\langle n \rangle = | \alpha |^2 $ that constitutes the pointer to the field amplitude, for any value of $ \alpha $. In addition, work is done by the field on the charges involved in the detection--this is quantified by  (9). 
  
  This brings us to the crucial point: the average number of photons absorbed by a suitable field detection system must correspond to  $ | \alpha |^2 $  in order to make possible the correspondence between the quantum level of the field and its classical counterpart. In particular, the energy $\bf{E}^2$ in the classical electric field component is proportional to  $ | \alpha |^2 $ because of the Born Rule (as well as because of Maxwell's equations!). Therefore, if the Born Rule were violated in such a way as to result in any detectable deviation, the macroscopic field would deviate as well, meaning that the energy in the field would no longer be proportional to $\bf{E}^2$. This is turn would mean that Maxwell's equations would not hold, nor would Poynting's theorem. Without adherence to the Born Rule, the classical statement of electromagnetic energy conservation, i.e., Poynting's theorem, would not apply to the processes taking place. Of course, energy would be conserved at the microscopic level, but if one considered a macroscopic volume V, one could find that the work done on charges in V, as given by (9), failed to satisfy Poynting's theorem (7) given (8), the latter being a straightforward consequence of Maxwell's equations.
  
 The more systematic and pronounced the Born Rule violation, the more detectable would be the apparent energy conservation violation. Although the result found herein is in terms of the electromagnetic field, the argument is not limited to massless fields; it applies to any field subject to energy conservation and coherent states.\footnote{Moreover, you could also lose phase information if the detected (absorbed) number of photons deviated significantly from $\langle n \rangle$. That is yet another obstacle to the possibility of Born Rule violation, which we will not go into here.}
 
 One might ask: isn't this just a matter of definition? I.e., isn't the amplitude of the coherent state simply defined to be proportional to the electric field amplitude? Similarly, isn't the electric field amplitude defined as being proportional to the field quadrature $X$--and is that why these observations tautologically follow? I would argue that these definitions are not arbitrary, but are forced on us by the empirical phenomena. The empirical success of quantum electrodynamics depends on the identification of the field quadratures with the electric and magnetic fields. Thus, rather than being an arbitrary definition, this relationship is a natural correspondence between the theory and observed phenomena. Regarding coherent states, if one prepared a different kind of quantum state, in which this relationship did not hold, that state would detectably not function as a support for the classical field. Since we know that the world is not really classical, at all levels, but is described at the microscopic level by quantum theory, it follows that \textit{some} quantum state of light must be the support of the phenomenal classical fields around us (e.g. in our radios, microwave ovens, televisions, etc.).  The only state that can do the job is the coherent state, and in order to do it, it must have an amplitude $\alpha \sim  E$. So apparently, nature creates her macroscopic electromagnetic fields out of coherent quantum states, and their amplitudes $\alpha$ possess this physical relationship to the classical field amplitude. No other definition works.
 
We might add in this connection that R. Serway \cite{Serway} makes use of this relationship between the classical field and the quantum level to motivate the Born Rule, by using the relationship between energy (as intensity) and the probability of photon detection. Thus this author is not the first to observe that the correspondence between quantum states and macroscopic fields is enforced by the Born Rule.\footnote{I am indebted to an anonymous referee for pointing this out.}  Serway uses a heuristic argument that does not make explicit use of coherent states; he merely shows that the amplitude of whatever quantum state supports the field must be proportional to $E$. However, a Fock state with definite photon number, e.g., $| n \rangle$,  cannot support the classical field, since its expectation value for the electric field operator is zero. Only coherent states will do. 
    
\section{Conclusions}  
  
The correspondence of the classical field amplitude $E $ with the amplitude $ \alpha $ of the quantum coherent state translates into a requirement that the Born Rule must be obeyed in order for Maxwell's Equations to hold and for the classical statement of energy conservation to be preserved.  This result has implications for some interpretations of quantum theory which allow for violations of the Born Rule and/or
the idea that the Born Rule is not a fundamental law of Nature. One such interpretation is the so-called Bohmian theory\cite{Bohm}.  According to the Bohmian theory, the Born Rule reflects the ``equilibrium distribution'' of particle positions, where the material particle is assumed to have a well-defined position at all times (whether measured or not) and to be guided by its associated wave function. As Valentini has emphasized, the Bohmian theory allows for deviation of the particles from their ``equilibrium'' condition. He states that according to the Bohmian theory,  ``...[quantum] noise is not fundamental but merely a property of an equilibrium state in which the universe happens to be at the present time. It is suggested that 'non-quantum' or nonequilibrium matter might exist today in the form of relic particles from the early universe.''(\cite{Valentini}, Abstract). However, the result obtained herein indicates that in the event of such a deviation of particles from their ``equilibrium condition'', classical continuity equations such as Poynting's Theorem in general would not hold (nor would the basic field equations). Thus, postulating that the Born Rule corresponds only to a contingent distribution of matter has significant consequences in terms of the applicability of conservation laws and field equations generally assumed to be fundamental; i.e., not contingent on any particular distribution of matter.

\section{Acknowledgment}

I am grateful to an anonymous referee for helpful suggestions.

\end{document}